\documentclass[twocolumn,noshowpacs,preprintnumbers,amsmath,amssymb]{revtex4}

\usepackage{verbatim}

\usepackage{graphicx}
\usepackage{dcolumn}
\usepackage{bm}

\begin{document}

\title{Thermodynamic calculations on the catalytic growth of multiwall carbon nanotubes}
\author{Christian Klinke}
\email{cklinke@us.ibm.com}
\altaffiliation{Present address: IBM Watson Research Center, 1101 Kitchawan Road, Yorktown Heights, NY 10598, USA.}
\affiliation{Institut de Physique des Nanostructures, Ecole Polytechnique F\'ed\'erale de Lausanne,\\
CH - 1015 Lausanne, Switzerland}

\author{Jean-Marc Bonard}
\altaffiliation{Present address: Rolex S.A., 3-7 Rue Francois-Dussaud, 1211 Geneva 24, Switzerland.}
\affiliation{Institut de Physique des Nanostructures, Ecole Polytechnique F\'ed\'erale de Lausanne,\\
CH - 1015 Lausanne, Switzerland }

\author{Klaus Kern}
\affiliation{Institut de Physique des Nanostructures, Ecole Polytechnique F\'ed\'erale de Lausanne,\\
CH - 1015 Lausanne, Switzerland \\ \textnormal{and} \\ Max-Planck-Institut f\"ur Festk\"orperforschung, D - 70569 Stuttgart, Germany \\}

\begin{abstract} 

We have developed a thermodynamic model of the catalytic growth of
multiwall carbon nanotubes from hydrocarbon precursors at elevated
temperature. Using this model we have computed the heat
distribution, and carbon concentration in the catalyst.
Calculations delivered a analytical formula for the growth time
and growth rate. We find that the growth is mainly driven by a
concentration gradient within the catalyst, rather than a
temperature gradient.

\end{abstract}

\maketitle

\section{Introduction}

Since the discovery of carbon nanotubes in 1991 by
Iijima~\cite{IIJIMA} there has been significant progress in their
improving synthesis~\cite{KERN,DAI} and developing technological
applications~\cite{CHATELAIN4,OOSTERKAMP,OKUYAMAB,ZHOU}. However,
the growth mechanism of carbon nanotubes remains poorly
understood. Indeed, continuted optimization of carbon nanotube
synthesis will only be possible if the growth mechanisms are
understood quantitatively. Theoretical studies had been performed
on the growth mechanism of carbon nanotubes on the atomic
scale~\cite{CAR,YAKOBSON}, or on the role of the catalyst during
the construction of nanotubes~\cite{FROUDAKIS,GERSTEN}. Most of
them consider the growth of singlewall
nanotubes~\cite{MAITI,PENNYCOOK}. Our approach discusses the
experimental and theoretical facts relevant for the catalytic
growth of multiwall nanotubes by chemical vapor deposition (CVD),
and takes them as basis for a more macroscopic, thermodynamical
model. It allows considering the whole system of catalyst
particle, nanotube and substrate, and calculating the growth time
and growth rate in a typical CVD process. The model is inspired by
earlier works~\cite{KERN4,KERN2,KLINKE3} and we extend the model
to take into account the geometry and thermal properties of the
catalyst. We use a finite element method (FEM) to compute the heat
generation and distribution, and the carbon migration in the
catalyst. It was found that the growth is mainly driven by a
concentration gradient as opposed to a thermal gradient, while the
process temperature plays a key role in terms of activating
diffusion. Considering the catalytic reactions of acetylene on
iron facets one can draw conclusions on the dependence of the
growth on the partial pressure. Furthermore, a mechanism for the
cessation of the growth is discussed. The calculations and
simulations are demonstrated here exemplarily for the nanotube
growth at 650$^{\circ}$C using iron as catalyst but may easily be
adapted to different conditions. This article is an attempt to
understand the nanotube growth with classical methods.

\section{Supposed growth mechanism}
\label{C-SUPPGROWTH}

\begin{figure}[!h]
\begin{center}
\includegraphics[width=0.45\textwidth]{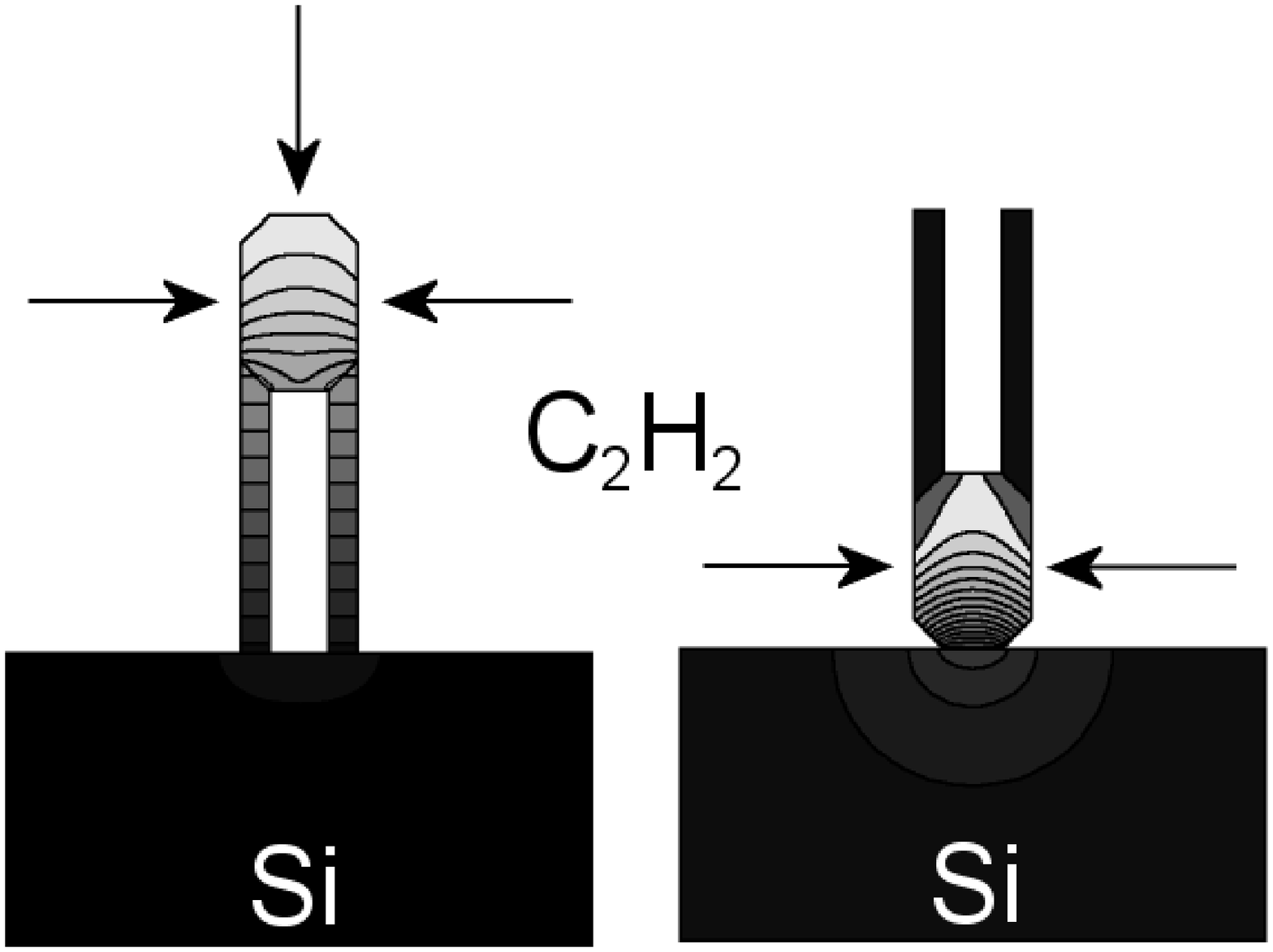}
\caption{\it Model for the suggested growth mechanism of catalytically grown carbon nanotubes and FEM simulations: Heat flow generated by the decomposition of acetylene at certain facets of the iron particle assuming a constant temperature at the silicon sample. Left: particle-on-top, right: particle-on-bottom setting. Arbitrary units for all dimensions (light: high temperature, dark: low temperature).}
\label{P-SIM-HEAT}
\end{center}
\end{figure}

It is widely believed that the mechanism of the catalytic growth of carbon nanotubes is similar to the one described by Kanzow et al.~\cite{DING}. Acetylene is thermally stable at temperatures below 800$^{\circ}$C and can be dissociated only catalytically, in the case discussed here, on the small metal (oxide) particles present on the substrate (Fig.~\ref{P-SIM-HEAT}). In a first phase the acetylene reduces the metal oxide particles to pure metal: Fe$_{2}$O$_{3}$ + 3C$_{2}$H$_{2}$ $\rightarrow$ 2Fe + 6C + 3H$_{2}$O, whereas the iron remains on the substrate surface, the carbon diffuses into the metal and the water evaporates. The further catalytic dissociation of acetylene takes presumably place at facets of well-defined crystallographic orientation~\cite{COULON}. The resulting hydrogen H$_{2}$ is removed by the gas flow while the carbon is dissolved in the catalyst. For unsaturated hydrocarbons this process is highly exothermic. When the particle is saturated with carbon, the carbon leaves the particle at another, less reactive surface of the particle. This process is endothermic. The resulting density gradient of carbon in the particle leads to diffusion of carbon through the particle. In order to avoid dangling bonds, the carbon atoms assemble in an $sp^2$ structure at a less reactive facet of the particle, ultimately leading to nanotube formation.

The simple model presented in Fig.~\ref{P-SIM-HEAT} describes the growth with a
catalyst particle at either the top or bottom of the tube. In principle both cases work in the same way, but in the latter one the particle adheres more firmly to the substrate surface than in the former. There must be free particle facets that are exposed to the gas for the growth to proceed. In the second case the acetylene diffuses from the sides into the particle and the nanotube is constructed from the bottom up, whereas in the first case the gas diffuses from the sides and top into the particle.

It was noted in~\cite{KERN2} that the nanotube growth did not begin immediately after the introduction of the hydrocarbon gas in the reactor, but that some carborized spots appear before rapid nanotube growth occurs. This suggests that a certain quantity of carbon must be dissolved in the catalyst before the nanotube growth can start. In addition, time is required for the oxided catalyst to be reduced.

\section{Calculations}

\subsection{Preconditions}
\label{C-PRECONDITIONS}

The catalytic reaction C$_{2}$H$_{2}$ $\stackrel{\mathrm{Fe}}{\rightarrow}$ 2C$_{\mathrm{graphitic}}$ + H$_{2}$ is highly exothermic. At 650$^{\circ}$C this reaction frees an energy of about 262.8~kJ/mol~\cite{NIST}. The two carbon atoms diffuse at a reactive facet into the catalyst particle and the hydrogen is taken away by the gas flow. The carbon diffuses through the particle to another less reactive facet where the carbon concentration is smaller and the temperature is lower. Similar models have been suggested by other authors \cite{DING,BAKER,WAITE,KANDA,KANDA2}.

The crystalline properties and the availability of defined facets are crucial points in the growth of carbon nanotubes. In an extensive study on catalytic particles on top of carbon fibers prepared by CO decomposition, Audier et al.~\cite{COULON} found that there are relations between the crystallographic structure of the catalyst particles and the attached nanotubes. In the case of a bcc structure of the catalyst particle, the particle is a single crystal with a [100] axis parallel to the axis of the fiber, and the basal facets of the truncated cone, which appeared free of carbon, are (100) facets. Anderson et al.~\cite{MEHANDRU} determined theoretically different activities of decomposition of acetylene on iron facets. Hung et al.~\cite{BERNASEK} mention that a complete decomposition of acetylene takes place at the Fe(100) facets with Fe(bcc), whereas molecular desorption was observed at the Fe(110) and Fe(111) facets. This may be due to differences in  surface roughness.

The crystalline character of the catalyst particles under our typical experimental conditions was proven by electron diffraction measurements and in situ real-space TEM images~\cite{KLINKE2}. At temperatures up to 1000$^{\circ}$C the catalyst particles are solid but a high material mobility and migration was observed during in situ heating of the catalyst. It has already been shown in~\cite{KLINKE3} that the diameter of the nanotube is determined by the size of the catalyst particle.

Lee et al.~\cite{CHO} found experimentally that the length of the nanotubes increases linearly with time. This suggests that the growth is a steady state process. Standard continuum modeling can then be used to calculate the heat flow and carbon diffusion in the catalyst. We start to calculate the carbon flow through the catalyst particle assuming a steady state. The non-steady state will be discussed later in the paper.

\subsection{Heat and particle diffusion}

The heat flow will be calculated with the particle flow (carbon atoms):

\begin{equation}\label{EQ-22}
\mathbf{j}_{\mathrm{p}} = -D \nabla c \qquad \textrm{Fick's First Law}
\end{equation}

where $D$ is the diffusion constant, $c$ the concentration, and $D$ is
given by

\begin{equation}\label{23}
D = D_{\mathrm{o}} \cdot \exp [-E_{\mathrm{a}} / kT] \qquad \textrm{Arrhenius equation}
\end{equation}

where $E_{a}$ is the activation energy and $D_{\mathrm{o}}$ the diffusion factor.\\

We carry out the calculation exemplarily with carbon in Fe(bcc)~\cite{CRC}, $D_{\mathrm{o}} = 2.2~\mathrm{cm^{2}/s}$, and $E_{\mathrm{a}} = 1.27~\mathrm{eV}$ (the case of carbon in Fe(fcc) is discussed later). Thus,
\begin{equation}\label{EQ-24}
D = 2.53 \cdot 10^{-11} ~\mathrm{\frac{m^{2}}{s}} \quad \textrm{at} ~923~\mathrm{K}~(650^{\circ}\mathrm{C})
\end{equation}

Following the iron-carbon-diagram, the maximal solubility of carbon in iron at 650$^{\circ}$C is $S = 65$~ppm(weight)~\cite{MASSALSKI}. Exceeding this amoung leads to the formation of iron carbide Fe$_{3}$C. This limit determines the maximal concentration gradient $\nabla c$. With Eq.~(\ref{EQ-22}),

\begin{equation}\label{26}
\arrowvert \nabla c \arrowvert  = \frac{S}{d_{\mathrm{diff}}} \cdot \frac{m_{\mathrm{mol}}[\mathrm{Fe}]}{m_{\mathrm{mol}}[\mathrm{C}]} \cdot \frac{1}{V_{\mathrm{mol}}[\mathrm{Fe}]}
\end{equation}

the diffusion distance $d_{\mathrm{diff}}$ (e.g. $d_{\mathrm{diff}} \simeq \frac{1}{2} d_{\mathrm{particle}}$), $m_{\mathrm{mol}}$[Fe] = $55.8$~g/mol,~$m_{\mathrm{mol}}$[C] = $12.0$~g/mol, and $V_{\mathrm{mol}}$[Fe] = $7.09 \cdot 10^{-6}~$m$^{3}$/mol, one obtains

\begin{equation}\label{EQ-27}
\arrowvert \nabla c \arrowvert = 42.63~\mathrm{\frac{mol}{m^{3}}} \cdot \frac{1}{d_{\mathrm{diff}}}
\end{equation}

With Eq.~(\ref{EQ-22}) and~(\ref{EQ-24}):

\begin{equation}\label{EQ-28}
j_{\mathrm{p,mol}}(\mathrm{bcc}) = 1.079 \cdot 10^{-9}~\mathrm{\frac{mol}{m \cdot s}} \cdot \frac{1}{d_{\mathrm{diff}}} \\
\end{equation}

\begin{displaymath}
\hookrightarrow \qquad j_{\mathrm{p,N}}(\mathrm{bcc}) = 6.499 \cdot 10^{14}~\mathrm{\frac{particle}{m\cdot s}} \cdot
\frac{1}{d_{\mathrm{diff}}}
\end{displaymath}

Since $C_{2}H_{2} \Rightarrow 2C + H_{2}$, we get the maximal heat flow through one facet with

\begin{equation}\label{EQ-29}
j_{\mathrm{q}} = j_{\mathrm{p}} \cdot 262.8~\mathrm{\frac{kJ}{mol}} \cdot \frac{1}{2}
\end{equation}

\begin{displaymath}
j_{\mathrm{q}}(\mathrm{bcc}) = 1.418 \cdot 10^{-4}~\mathrm{\frac{J}{m \cdot s}} \cdot \frac{1}{d_{\mathrm{diff}}}
\end{displaymath}

Based on the results of many experiments we can define a typical
nanotube: a hollow cylinder with a length of $l_{\mathrm{nt}} = 5~\mu$m, an inner diameter of $d_{\mathrm{in}} = 10$~nm and an outer diameter of $d_{\mathrm{out}} = 20$~nm $= d_{\mathrm{particle}}$ (compare with~\cite{KERN4}). It has the volume of

\begin{equation}\label{EQ-31}
V_{\mathrm{nt}} = \frac{\pi \cdot l_{\mathrm{nt}}}{4} (d_{\mathrm{out}}^{2} - d_{\mathrm{in}}^{2}) = 1.178 \cdot 10^{-21}~\mathrm{m^{3}}
\end{equation}

\begin{displaymath}
\stackrel{\wedge}{=} ~ 2.572 \cdot 10^{-16}~\mathrm{mol}~\stackrel{\wedge}{=} ~ 3.087 \cdot
10^{-18}~\mathrm{kg}
\end{displaymath}

\begin{equation}\label{EQ-32}
\textrm{with} \qquad V_{\mathrm{mol}} [\mathrm{C}] = 4.58 \cdot 10^{-6}~\mathrm{\frac{m^{3}}{mol}}
\end{equation}

\begin{displaymath}
\textrm{and} \quad m_{\mathrm{mol}} [\mathrm{C}] = 12.0~\mathrm{g} = 0.012~\mathrm{kg}
\end{displaymath}

Then the total of converted energy would be

\begin{equation}\label{33}
\Delta Q_{\mathrm{total}} = 262.8~\mathrm{\frac{kJ}{mol}} \cdot 2.572 \cdot 10^{-16}~\mathrm{mol} \cdot \frac{1}{2} = 3.38
\cdot 10^{-11}~\mathrm{J}
\end{equation}

And if all the heat was accumulated just in the catalyst particle and the nanotube, without a transfer to another reservoir, the nanotube would be heated up by

\begin{equation}\label{EQ-34}
\Delta T = \frac{\Delta Q_{\mathrm{total}}}{c_{\mathrm{q}}[\mathrm{Fe}] \cdot m[\mathrm{Fe}] + c_{\mathrm{q}}[\mathrm{C}] \cdot m[\mathrm{C}]} = 6834~\mathrm{K}
\end{equation}

With the heat capacity $c_{\mathrm{q}}[$Fe$] = 449,0~\mathrm{\frac{J}{kg \cdot K}}$, $m[$Fe$] = \frac{\pi}{6} \cdot (20~\mathrm{nm})^{3} \cdot \rho [$Fe$]= 3.292 \cdot 10^{-20}$~kg (e.g. sphere as catalyst particle, diameter: 20~nm), $c_{\mathrm{q}}[$C$] = 710,0~\mathrm{\frac{J}{kg \cdot K}}$, $m[$C$] = 6,945 \cdot 10^{-18}$~kg. \\

This seems to be a very high temperature increase. But one has to bear in mind that we have assumed an isolated nanotube, without any contact to the environment. The actual temperature rise will be much lower due to heat conduction by the substrate, as is discussed below.

\subsection{Growth time and rate}
\label{C-FURTHERCALC}

It is possible to estimate the growth time $t_{\mathrm{growth}}$ for a nanotube. With Eq.~(\ref{EQ-28}) and the number of moles of the
nanotube $n_{\mathrm{nt}}$, this is

\begin{equation}\label{EQ-35}
t_{\mathrm{growth}} = \frac{n_{\mathrm{nt}}}{j_{\mathrm{p}} \cdot \frac{1}{2}A_{\mathrm{particle}}}
\end{equation}

For the standard nanotube ($n_{\mathrm{nt}} = 2.572 \cdot 10^{-16}~\mathrm{mol}$), we get $t_{\mathrm{growth}}(\mathrm{bcc}) = 3.8~\mathrm{s}$. Using the parameters for carbon in Fe(fcc)~\cite{CRC} ($D_{\mathrm{o}} = 0.15~\mathrm{cm^{2}/s}$, $E_{\mathrm{a}} = 1.47~\mathrm{eV}$, thus $D = 1.33 \cdot 10^{-13} ~\mathrm{m^{2}/s}$ at 923~K) yields $t_{\mathrm{growth}}(\mathrm{fcc}) = 722~\mathrm{s}$. Analogously, the growth rate is $v_{\mathrm{growth}} = l_{\mathrm{nt}}/t_{\mathrm{growth}}$, yielding $v_{\mathrm{growth}}(\mathrm{bcc}) = 1.3~\mathrm{\mu m/s}$ and $v_{\mathrm{growth}}(\mathrm{fcc}) = 6.9~\mathrm{nm/s}$ for bcc and fcc iron, resp. In the following we regard just Fe(bcc) because experimental results indicate growth rates in the order of 0.1 - 5~$\mu$m/s
assuming typical CVD conditions.

In order to evaluate the influence of the temperature on the growth time, Eq.~(\ref{EQ-35}) can be written (by using Eq.~(\ref{EQ-22}), (\ref{EQ-31}) and~(\ref{EQ-32})) as

\begin{equation}\label{37}
t_{\mathrm{growth}} = \frac{l_{\mathrm{nt}} \cdot (d_{\mathrm{out}}^{2} - d_{\mathrm{in}}^{2})}{2 \cdot V_{\mathrm{mol}} [\mathrm{C}] \cdot d_{\mathrm{particle}}^{2} \cdot D \cdot \arrowvert \nabla c\arrowvert }
\end{equation}

The solubility S of carbon in iron as function of temperature in the range from 500 to 740$^{\circ}$C shows an essentially exponential dependence on temperature
 $S = S_{\mathrm{o}} \cdot \exp[- \beta / T]$, with $S_{\mathrm{o}} = 30.593$ and $\beta = 12028.62$~K. $\beta$ can be expressed in units of an energy as $E_{\mathrm{S}} =\beta k = 1.036$~eV. The temperature dependance of the diffusion coefficient $D$ is descripted by Eq.~(\ref{23}). We assume that only the solubility and diffusion constant are temperature dependent. With Eq.~(\ref{26}), this leads to

\begin{equation}\label{38c}
t_{\mathrm{growth}} = \frac{\rho [\mathrm{C}]}{\rho [\mathrm{Fe}]} \cdot \frac{l_{\mathrm{nt}} \cdot d_{\mathrm{diff}} \cdot (d_{\mathrm{out}}^{2} - d_{\mathrm{in}}^{2})}{2 d_{\mathrm{particle}}^{2} \cdot D_{\mathrm{o}} \cdot S_{\mathrm{o}} } \cdot \exp [\frac{E_{\mathrm{a}} + E_{\mathrm{S}}}{kT}]
\end{equation}

With $d_{\mathrm{nt}} = d_{\mathrm{particle}} = d_{\mathrm{out}} = 2 \cdot d_{\mathrm{in}} = 2\cdot d_{\mathrm{diff}}$, the growth rate $v_{\mathrm{growth}} = l_{\mathrm{nt}} / t_{\mathrm{growth}}$ turns then out to be:

\begin{equation}\label{EQ-41a}
v_{\mathrm{growth}} = \frac{16}{3} \cdot \frac{\rho [\mathrm{Fe}]}{\rho [\mathrm{C}]} \cdot \frac{D_{\mathrm{o}} \cdot S_{\mathrm{o}}}{d_{\mathrm{nt}}} \cdot \exp [-\frac{E_{\mathrm{a}} + E_{\mathrm{S}}}{kT}]
\end{equation}

\begin{figure}[!h]
\begin{center}
\includegraphics[width=0.45\textwidth]{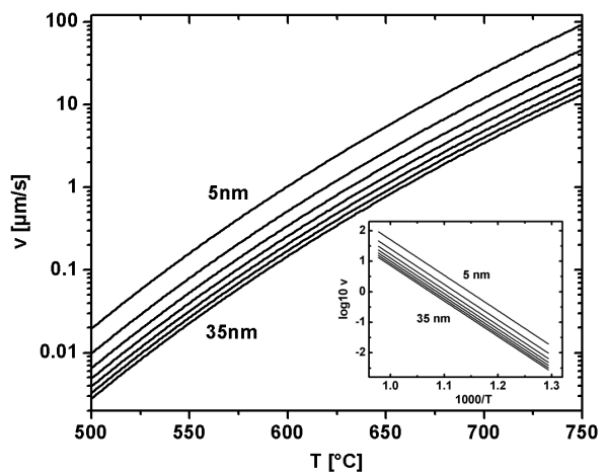}
\caption{\it Calculated growth rate as function of deposition temperature (500-750$^{\circ}$C) and nanotube diameter (5-35~nm in 5~nm increments). The inset shows the curves as Arrhenius plot.}
\label{P-CALCGROWTH}
\end{center}
\end{figure}

The growth rate $v_{\mathrm{growth}}$ is proportional to $\exp [-(E_{\mathrm{a}} + E_{\mathrm{S}})/kT]$ (which obviously has a strong temperature dependence) and to $1/d_{\mathrm{nt}}$. This is in accordance with the experimental fingings of Lee et al.~\cite{CHO} for carbon nanotubes, and similar to the result Baker~\cite{BAKER} found experimentally for carbon filaments $v_{\mathrm{growth}} \propto 1 / \sqrt{d_{\mathrm{particle}}}$. The $1/d_{\mathrm{nt}}$ behavior is demonstrated in Fig.~\ref{P-CALCGROWTH}. The calculated growth rate is shown as function of the deposition temperature in the range between 500 and 750$^{\circ}$C and the nanotube diameter (5-35~nm in 5~nm increments).

In order to incorporate the effect of catalytic reactions at the surface of the
catalyst particle, we start with

\begin{equation}\label{42}
v_{\mathrm{growth,N}} = j_{\mathrm{p,N}} \cdot A_{\mathrm{creation}} = j_{\mathrm{p,N}} \cdot \frac{1}{2}A_{\mathrm{particle}}
\end{equation}

where $A_{\mathrm{creation}}$ is the surface of nanotube generation. For the standard nanotube this is $v_{\mathrm{growth,N}} = 4.083 \cdot 10^{7}~\mathrm{particle / s}$. This may also be the rate of carbon atoms converted at the facets. Thus the reaction rate is $v_{\mathrm{reaction}} = \frac{1}{2} \cdot v_{\mathrm{growth,N}} = 2.042 \cdot 10^{7}~\mathrm{reaction / s}$ ($C_{2}H_{2} \Rightarrow \underline{\underline{2C}} + H_{2}$). The reaction flow is then $j_{\mathrm{reaction}} = v_{\mathrm{reaction}} / (\frac{1}{2}A_{\mathrm{particle}}) = \frac{1}{2} \cdot j_{\mathrm{p,N}} = 3.25 \cdot 10^{22} ~ \mathrm{reaction / (m^{2}
\cdot s)}$.

If one assumes that the carbon stock is an ideal gas,
the impingement rate of gas particles on the catalyst is

\begin{equation}\label{46}
j_{\mathrm{impact}} = \frac{p}{\sqrt{2 \pi m k T}}
\end{equation}

where $m$ is the mass of the gas molecule (here: $m$[C$_{2}$H$_{2}$] $= 4.324 \cdot 10^{-26}$~kg, $T = 923$~K) and $p$ the pressure. At e.g. 20~mbar = 2000~Pa (standard experiment) we have then an impact rate of $j_{\mathrm{impact}} = 3.4 \cdot 10^{25}~\mathrm{hits / (m^{2} \cdot s)}$. Compared with the
reaction rate $j_{\mathrm{reaction}} = 3.25 \cdot 10^{22}~\mathrm{reaction / (m^{2} \cdot s)}$ this means that the surface is always saturated. A reduction of the growth rate should then occur at pressures under $p = j_{\mathrm{reaction}} \cdot \sqrt{2 \pi m k T} = 1.913~\mathrm{Pa} = 1.913 \cdot
10^{-2}~\mathrm{mbar}$.  A sticking coefficient less than 1 will influence these results.
Note that,
in principle, the variation of growth rate with pressure can be used to control the length of
the nanotubes.

\section{Simulations}
\label{C-GROWTHSIM}

We used two-dimensional finite element method (FEM) simulations to compute the temperature distribution in the particle, the nanotube and the substrate. The carbon concentration in the catalyst particle was simulated as well. Specifically for the temperature distribution we compute $\partial T / \partial t - \kappa \triangle T = 0$ and in the stationary case $- \kappa \triangle T = 0$. The boundary conditions for a constant heat flow through the facet is $\nabla T = -\mathbf{j}_{\mathrm{q}} / \lambda ~ | _{\mathrm{facet}}$ (with the heat conductivity $\lambda $)  and for a constant temperature at the facet $T = T_{\mathrm{b}} ~ | _{\mathrm{facet}}$. Accordingly for the distribution of the carbon concentration we compute $\partial c / \partial t - D \triangle c = 0$ and in the stationary case $- D \triangle c = 0$. The boundary conditions for a constant particle flow through the facet is $\nabla c = -\mathbf{j}_{\mathrm{p}} / D ~ | _{\mathrm{facet}}$ (with the diffusion constant $D $) and for a constant carbon concentration at the facet $T = c_{\mathrm{b}} ~ | _{\mathrm{facet}}$.

At the boundary we assume a heat penetration of $\mathbf{j}_{\mathrm{q}}$ at the horizontal and vertical facets of a model catalyst particle (bcc(100)-like facets, see Section~\ref{C-PRECONDITIONS}) and constant temperature of $T_{\mathrm{b}}$ at the three outer sides of the silicon substrate (Fig.~\ref{P-SIM-HEAT}). A distribution of temperature rise is then obtained for the defined geometry. The shape of the model particle is chosen to approximate those observed in experiments (e.g.~\cite{KERN4}). The particle can be on top of the nanotube (pushed up by the nanotube) or sticking to the substrate surface (while pushing up the nanotube). The silicon substrate has a length of 10~$\mu$m and height of 5~$\mu$m, which seems to be sufficient (a further enlargement did not cause a change in temperature in the simulations).

For the heat conductivity of the nanotube, the value for graphite in the direction parallel to the graphitic layers is assumed, rather than the (much higher) value for perfect single-wall nanotubes ($\lambda = 2980.0~\mathrm{W / (m \cdot K)}$~\cite{GODDARD}). We therefore obtain an upper limit for the temperature rise. For iron, $\lambda$ is taken at the same temperature as the value for silicon and graphite (373.2~K) although a value at 900~K
has been reported (for iron only).

\begin{figure}[!h]
\begin{center}
\includegraphics[width=0.45\textwidth]{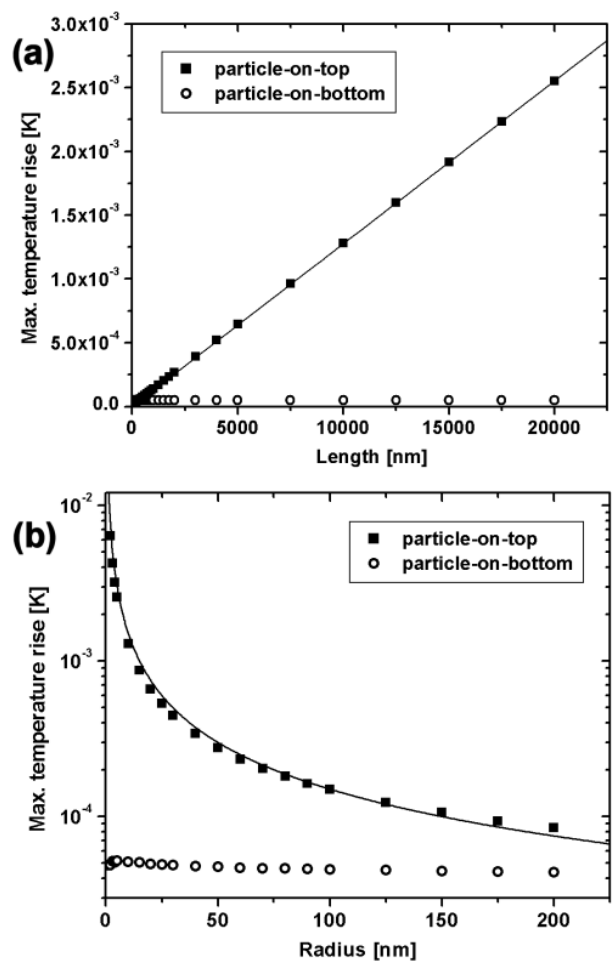}
\caption{\it Dependence of the maximal temperature rise on (a) the nanotube length and (b) the particle radius using the particle-on-top and the particle-on-bottom setting. The simulated values for the particle-on-top setting fit well with a linear behavior in (a) and with a 1/x behavior in (b).  The solid lines are according guides to the eye.}
\label{P-DIAGRAM}
\end{center}
\end{figure}

The particle-on-top setting is highly parameter dependent. The dependence of the maximum temperature in the particle on the nanotube length
is approximately linear $T_{\mathrm{max}} \sim l_{\mathrm{nt}}$ (Fig.~\ref{P-DIAGRAM}a), while the dependence on  nanotube radius varies as $1/x$ law: $T_{\mathrm{max}} \sim 1/r_{\mathrm{nt}}$ (Fig.~\ref{P-DIAGRAM}b). The standard nanotube ($l_{\mathrm{nt}} = 5$~$\mu$m, $r_{\mathrm{nt}} = 10$~nm, $d_{\mathrm{diff}} =$ 20~nm) reaches a temperature rise of $\Delta T_{\mathrm{top}} = 6.474 \cdot 10^{-4}$~K with the particle on top (Fig.~\ref{P-SIM-HEAT}). For the standard tube 90~\% of the final temperature rise is reached already after 1.65~$\mu$s. In contrast, in the particle-on-bottom configuration the temperature rise is independent of the tube geometry. In this configuration the standard setting reaches a temperature rise of $\Delta T_{\mathrm{bottom}} = 5.036 \cdot 10^{-5}$~K.

According simulation have been undertaken also for the distribution of the carbon concentration in the catalyst particle in the non-steady state mode. At the highly reactive facets we assume a flow corresponding to Eq.~\ref{EQ-28}. On the less reactive facets a carbon concentration of 0~ppm is asssumed. It turned out that, for the standard nanotube, it takes about 100~$\mu$s to reach 90~\% of the final carbon concentration at one facet. This satisfies the assumption of steady state conditions in the catalyst particle during the growth, since the time for the complete growth lies in the range of seconds.

As the maximal temperature rise in the particle is very low, the diffusion of carbon through the particle can be considered to be driven solely by the concentration gradient. It is assumed that the concentration of carbon is 65~ppm(weight) on the facets where the catalytic decomposition takes place and 0~ppm where the nanotube is assembled. The diffusion through the particle is hence determined by Eq.~(\ref{EQ-22}).

\section{Discussion}

We have computed the growth rate of carbon nanotubes based on a new growth mechanism.
We show that for realistic tube/particle geometries the expected temperature rise is negligible.
Thus thermal radiation is also negligible and the carbon diffusion through the particle
is not thermally driven but is determined by the carbon concentration gradient.

The difference between the calculated temperature rise $\Delta T = 6834$~K if all the produced heat is
retained in the particle and the nanotube (Eq.~(\ref{EQ-34})) and the simulated $\Delta T = 6.474 \cdot 10^{-4}$~K is due to the thermal coupling to the substrate in the latter case and the high thermal conductivities of iron, graphite and silicon. The produced heat is distributed in the material very rapidly (diffusive flux through the iron particle, the nanotube, the silicon substrate and out of the considered volume). In the particle-on-bottom
scenario this flux is led away even more rapidly and the temperature rise is consequently smaller (diffusion direct into silicon substrate).

At first glance there seems to be no reason why the nanotubes should grow in one particular direction. But the silicon surface breaks the symmetry. The diffusion and the catalytic decomposition is favored at the top facets and the catalyst particle is pushed up by the growing nanotube. Only if the
particle is bonded too strongly to the silicon surface the nanotube grows in upward direction while the particle remains on the surface.

Under the considered conditions, growth by surface diffusion is unlikely because it cannot explain the growth of multiwall nanotubes (growth of several walls with the same velocity, diffusion of carbon through the already created walls). Hung et al.~\cite{BERNASEK} report that carbon diffuses into bulk iron starting at temperatures $T > 773$~K ($500^{\circ}$C). This means that multiwall carbon nanotubes cannot be created below this temperature. The mobility of carbon in iron will increase with the temperature since the diffusion coefficient $D$ is highly temperature dependent. This explains why the nanotube growth is a thermal process. A certain mobility of the carbon atoms in the catalyst particle is essential for the growth. According to Hung et al.~\cite{BERNASEK} C$_{2}$H$_{2}$ decomposes catalytically to pure C at temperatures T $>$ 400~K (127$^{\circ}$C). Thus there is a well defined window for the growth of pure nanotubes between $500^{\circ}$C and $800^{\circ}$C. At higher temperatures polycrystalline carbon adsorbes on the nanotube surface~\cite{KERN4}. In the frame of the experiments discussed in~\cite{KERN4,KERN2} the lowest growth temperature was 620$^{\circ}$C. For plasma-enhanced CVD lower deposition temperatures are reported: e.g. Choi et al. observed the experimental growth of nanotubes at 550$^{\circ}$C~\cite{KIM2}.

The cessation of the catalytic growth may be caused by the formation of amorphous carbon on the catalyst particle.
This can be generated catalytically or by condensation of decomposed hydrocarbons~\cite{KERN4}. Acetylene can be cracked at relative low temperatures due to the Boltzmann distribution of the thermal energy of the gas.
There are always some gas molecules with enough energy for the cracking. Of course, the proportion increases with higher temperatures and thus finally there will be more amorphous carbon. Hung et al.~\cite{BERNASEK} report a blocking of the reaction sites in case of high carbon coverage. Additionally an oversupply of carbon (more then 65~ppm(weight) on the reaction surface) can cause the formation of iron carbide Fe$_{3}$C. The diffusion of carbon through Fe$_{3}$C is very low ($D = 6 \cdot 10^{-16} \mathrm{m^{2} / s}$ at $650^{\circ}$C~\cite{SIMKOVICH}). If Fe$_{3}$C is generated on the reaction surfaces, this
can also stop the growth of the nanotubes. H$_{2}$ in the gas flow can etch the oversupply of amorphous carbon~\cite{AJAYAN} and extend the growth time and thus the nanotube length. Indeed, this prolongation of the nanotubes was observed experimentally (e.g. by Kim et al.~\cite{DAI}). A reduction of pressure may help as well, but this would also slow down the growth of nanotubes.

The calculations allow us to estimate the growth time and growth rate. It is questionable if continuum modeling is valid on the nanometer scale, but the calculated results correlate well with the experimental data. In Table~\ref{T-ACTIVATION}, some experimental data for the nanotube growth are listed and compared with calculated values. The calculated values for the growth rate fall quite well in the range of the experimental data of our own group, and the calculated growth time is at the lower limit of the experimental data. From other experimental groups just limited data are available, and their conditions can differ considerably from ours. Thus, the growth rate could be limited by factors not considered here (gas pressure/flow, homogeneity of temperature, catalyst preparation, etc.). Two examples for growth rates obtained by other groups are mentioned in Table~\ref{T-ACTIVATION}. Their values are about one order of magnitute lower than our results. It has to be considered that the calculated values are for ideal standard tubes. However, the diameter of the tubes in the experiments vary usually between 5 and 50~nm and the length between 2 and 10~$\mu$m and not all conditions of CVD processes are well known. Maiti et al.~\cite{BERNHOLC,BERNHOLC2} evaluated a catalyst free growth model by molecular dynamics and found a growth rate of 160~nm/s at 1000~K. This is substantially lower than what we found experimentally as well as with the calculations presented here. The reason for this might lie in the importance of the catalytic growth by the according particle. The catalytic decomposition and the accompanying diffusion seem to accelerate the growth considerably.

According to our calculations the growth rate is not influenced by a varying length but just by the diameter of the nanotube as $v_{\mathrm{growth}} \propto 1 / d_{\mathrm{nt}}$. Baker et al.~\cite{WAITE} found experimentally a dependence of $v_{\mathrm{growth}} \propto 1 / \sqrt{d_{\mathrm{nt}}}$ for carbon filaments. The difference between the $1 / \sqrt{d_{\mathrm{nt}}}$ dependence for filaments and the calculated dependence $1 / d_{\mathrm{nt}}$ for nanotubes may lie in the fact that carbon nanotubes are hollow graphite-like structures and the carbon filaments consist of monolithic amorphous carbon, which may result in a slower growth. However, Lee et al.~\cite{CHO} found experimentally the same $v_{\mathrm{growth}} \propto 1 / d_{\mathrm{nt}}$ dependence for carbon nanotubes. They also found that the growth time is linearly proportional to the nanotube length $t_{\mathrm{growth}} \propto l_{\mathrm{nt}}$. This is another hint that the growth is mainly a steady state process and thus the usage of Fick's First law is justified. Our model also reflects their finding that the growth rate increases superproportionally with temperature~\cite{LEE2}.

The discussed activation time (time when the catalyst is already exposed to acetylene but the nanotube growth has not yet started) was observed as well~\cite{KERN2}. This period might be due to the reduction of Fe$_{2}$O$_{3}$ to pure iron.

In order to directly compare the experimental data with the calculations the growth needs to be observed in situ to determine the growth time of one individual nanotubes and to determine at the same time the length of this tube and its diameter. For carbon filaments this has was done as early as 1972 by Baker et al.~\cite{WAITE}. But detailed studies for carbon nanotubes are still lacking because in situ growth measurements on nanotubes are difficult to perform.

We have shown that the growth rate is a function of the applied partial pressure, and that effects should be evident below a pressure of about 2 $\cdot$ 10$^{-2}$ mbar. In~\cite{CHATELAIN5} and~\cite{CHATELAIN6} the influence of the partial pressure on the growth velocity is addressed, e.g. in ~\cite{CHATELAIN5} for the pressures 10$^{-4}$, 10$^{-3}$ and 10$^{-2}$~mbar growth rates of 1.5, 3.7 and 4.7~$\mu$m/s are
measured.

While our calculations are necessarily approximate, they give clear quantitative picture of the nanotube generation process. Naturally, a different tube/catalyst geometry will lead to different values of  $v_{\mathrm{growth}}$ and $t_{\mathrm{growth}}$. The calculations can easily be repeated with other material parameters (e.g. for CH$_{4}$ as carbon source gas and nickel as catalyst particle).

\section{Conclusions}

We performed thermodynamic modeling of the catalytic generation of carbon nanotubes which allow us to estimate the growth time and growth rate. The conclusions of the model are supported by experimental results. We find that the growth is mainly driven by carbon concentration gradients in the catalyst particle, rather than by temperature gradients.

\section*{Acknowledgement}

The Swiss National Science Foundation (SNF) is acknowledged for the financial support.

\clearpage

\section*{Figures}

\clearpage

\section*{Tables}

\begin{table}[!h]
\begin{center}
\caption {\it Comparison of calculated values with available experimental data for growth time and growth rate.}
\label{T-ACTIVATION}
\vspace{0.5cm}
\begin{tabular}{|c|c|c|c|c|} \hline
\textbf{Source} & \textbf{Growth time [s]} & \textbf{Growth rate [$\mu$m/s]} & Temperature [$^{\circ}$C] & Diameter [nm]\\ \hline
Calculated (in text)    & 3.8       &  1.3      & 650   & 20        \\ \hline
Calculated (not shown)  & 0.4       &  12.1     & 700   & 10        \\ \hline
Ref.~\cite{KERN2}       & $<$ 60    & $>$ 0.16  & 720   & -         \\ \hline
Ref.~\cite{KLINKE3}     & $<$ 10    & $>$ 0.1   & 720   & 5         \\ \hline
Ref.~\cite{CHATELAIN5}  & 10        & 0.9 - 5.1 & 700   & 5 - 10    \\ \hline
Ref.~\cite{CHATELAIN6}  & 10 - 15   & 0.9 - 8.0 & 700   & 5 - 10    \\ \hline
Ref.~\cite{HOWE}        & -         & 0.1       & 650   & -         \\ \hline
Ref.~\cite{LEE}         & -         & 0.27      & 800   & 15 - 25   \\ \hline
\end{tabular}
\end{center}
\end{table}

\clearpage

\end{document}